\documentstyle[12pt]{article}

\def\half{1 \over 2}
\def\bp{\partial \!\!\!\partial}

\def\>{\rangle}
\def\<{\langle}

\def\beq{\begin{equation}}
\def\eeq{\end{equation}}
\def\bed{\begin{displaymath}}
\def\eed{\end{displaymath}}
\def\beqq{\begin{eqnarray}}
\def\eeqq{\end{eqnarray}}
\def\bedd{\begin{eqnarray*}}
\def\eedd{\end{eqnarray*}}

\advance\hoffset by -.35in
\begin{document}

\begin{flushright}
{\bf DIAS-STP 96-22\\Nov 1996}
\end{flushright}

\def\sqr#1#2{{\vcenter{\vbox{\hrule height.#2pt
\hbox{\vrule width.#2pt height#1pt \kern#1pt
\vrule width.#2pt}
\hrule height.#2pt}}}}
\def\square{\mathchoice\sqr34\sqr34\sqr{2.1}3\sqr{1.5}3}

\begin{center}\huge{\bf{Brief Resume of Seiberg-Witten Theory}}
\end{center}
\vspace{1cm}

\begin{center}
R. Flume\footnote{Volkswagen Foundation Fellow, on leave of absence 
from University of Bonn}, L.O'Raifeartaigh\footnote{Talk presented at the Inaugural Coference of the Asia  
Pacific Center for Theoretical Physics, Seoul, June 1996,  
the XXIst Conference on Group Theor. Methods in Physics, 
Goslar, Germany, July 1996 and 2nd SIMI Conference, Tbilisi, Sept. 
1996} and I.Sachs\footnote{Partially supported by Swiss Nationalfond for Scientific 
Research}\\
{\it Dublin Institute for Advanced Studies\\
School of Theoretical Physics\\
10 Burlington Road, Dublin 4, Ireland}
\end{center}
\vspace{1cm}

\def \rref#1#2{\item{[#1]}#2}
\def \tref#1{{\bf[#1]}}\def \mtxt#1{\quad\hbox{{#1}}\quad}
\def\half{1 \over 2}\def \lwd#1{\lower2pt\hbox{$\scriptstyle #1$}}
\def \parn {\par \noindent}
\def \bp {\partial \!\!\!\partial}
\overfullrule=0pt\rightline{ }
\noindent
{\bf 1. Introduction} 
\vskip 0.3truecm \noindent
An outstanding problem in non-abelian gauge theory has been to make 
reliable predictions about the (non-perturbative) strong coupling region. 
(Another interesting problem has ben to find solvable models in more 
than two dimensions). In a recent paper [1] Seiberg and Witten (SW) have 
shown that the simplest non-trivial $N=2$ supersymmetric theory 
provides at least a partial answer to these problems. First, 
they have shown that the local part of the effective Action is 
governed by a single analytic function $F$ of a complex variable. 
Second they have made an Ansatz for the $F$ that satisfies 
all the physical criteria and embodies electromagnetic duality,  
thus directly connecting the weak to the strong 
coupling regions.  The correctness of their Ansatz is supported by 
some direct instanton computations [2]. The purpose of this note is to 
give a resume of the SW theory in the simplest possible 
mathematical terms. 
\vskip 0.3truecm \noindent 
{\bf 2.} ${\bf N=2}$ {\bf Supersymmetry.}
\vskip 0.3truecm \noindent 
We begin by recalling the essentials [3] of the $N=2$ supersymmetry 
algebra and its Action. The algebra is
$$\{ Q_\alpha^i,\bar Q_\beta^k \}=\delta_{ik}\sigma^\mu_{\alpha 
\beta}P_\mu 
\qquad \{ Q_\alpha^i, Q_\beta^k \}=\epsilon_{ik}\epsilon_{\alpha 
\beta} Z  \eqno(2.1)$$ 
plus the hermitian conjugate of the second relation, where $i,k=1,2$ and $Z$ is a central charge. This algebra is 
realized on the simplest possible non-trivial supermultiplet, namely 
$$\Psi  \supset \{\phi,\psi,A_\mu; F,D  \} \eqno(2.2)$$
where $\phi$ is a complex scalar field, $\psi$ is a Dirac spinor 
$A_\mu$ is a gauge-field and $F$ and $D$ are complex and 
realdummy-fields respectively. This $N=2$ superfield actually 
consists of two $N=1$ superfields, namely 
$$\Phi \supset\{\phi,q,F\}   \quad \hbox{and} \quad 
V \supset \{A_\mu,f, D\}    \quad \hbox{or} \quad 
W_\alpha \supset  \{F_{\mu\nu},f,D\}  \eqno(2.3)$$
where $\phi$ and $V/W_\alpha$ are chiral and vector 
multiplets respectively, the $q$ 
\vskip 0.8truecm \noindent

\noindent and $f$ fields being Weyl spinors of opposite chirality.  Since 
$A_\mu$ belongs to the adjoint representation of the gauge group 
$G$ and all the fields belong to the same multiplet they must all 
belong to the adjoint representation of $G$. The simplest SW model 
is for $G=SU(2)$ and we shall concentrate on this case. 

\vskip 1truecm \noindent 
{\bf 3. }${\bf N=2}$ {\bf Super-Action.} 
\vskip 0.3truecm \noindent 
The superaction for the $N=2$ superfield just described is 
$${\cal A}=\hbox{Im Tr}\int d^4xd^2\theta_\alpha d^2\bar\theta_\beta 
\Bigl(\Psi\Bigr)^2  \eqno(3.1)$$ 
On expanding this in terms of the $N=1$ 
superfields it becomes $${\cal A}=\hbox{Im}\int d^4xd^2\theta_\alpha 
d^2\bar\theta_\beta \Bigl(\bar Ae^{-2g_oV}A\Bigr) +\tau_o\int 
d^4xd^2\theta_\alpha \Bigl( W_\alpha W_\alpha\Bigr)  \eqno(3.2)$$ 
where $$\tau_o={\theta_o \over 2\pi} +{4\pi i \over g_o^2}  
\eqno(3.3)$$ 
the parameter $g_o$ being the usual gauge-coupling 
constant and $\theta_o$ being the $QCD$-vacuum-angle (not to be 
confused with the usual supersymmetric Grassman variables). The 
exponential in the first term is just the supersymmetric 
generalization of the covariant derivative. Expanding (3.2) further 
in terms of conventional fields we obtain 
\beqq {\cal 
A}=\hbox{tr}\int d^4x &\Bigl\{{1 \over 2}(\phi^\dagger D^2\phi) 
+\bar\psi D\psi +g_o(\phi[\bar \psi,\gamma_5\psi]) 
+g_o^2[\phi^\dagger,\phi]^2  
\Bigr\} \nonumber\\ &+\hbox{tr}\int d^4x \Bigl\{{1 \over 
4g_o^2}F^{\mu\nu}F_{\mu\nu}+{\theta_o \over 32}\tilde 
F^{\mu\nu}F_{\mu\nu}   \Bigr\} \nonumber
\eeqq 
\vspace{-7ex}
$$\eqno(3.4)$$
This Action will 
be immediately recognized as the standard action for a 
Quark-Gluon-Higgs system in which all the fields are in the adjoint 
representation and the coupling constants are reduced to $g$ and 
$\theta$ by the supersymmetry. Thus it is not very exotic. Indeed it 
could be the $QCD$ Action except for the fact that the quarks are in 
the adjoint and presence of the scalar field. 
\vskip 0.3truecm\eject \noindent 
{\bf 4. }{\bf Text-Book Properties} 
\vskip 0.3truecm \noindent 
The Action (3.4) is actually so normal that it embodies 
all the properties of Quantum gauge Theory that have surfaced over 
the past thirty years and could even be used as a model to teach 
quantum gauge theory. It might be worthwhile to list these 
properties: 
\vskip 0.2truecm \noindent 1. 
It contains a gauge-field coupled to matter 
\vskip 0.2truecm \noindent 2. 
It is asymptotically free 
\vskip 0.2truecm \noindent 3. 
It is scale-invariant, but with a scale-anomaly 
\vskip 0.2truecm \noindent 4. 
It has spontaneous symmetry breaking 
\vskip 0.2truecm \noindent 5. 
It has central charges ($Z$ and $\bar Z$) 
\vskip 0.2truecm \noindent 6. 
It admits both instantons and monoples 
\vskip 0.2truecm \noindent 
Because of the supersymmetry it has some further special properties, 
whose significance will become clear later, namely, 
\vskip 0.2truecm \noindent 7. 
It not only has a Montonon-Olive mass formula [4] for 
gauge-fields and monopoles but generalizes that formula 
$$\hbox{from} \quad M=\vert v\vert \bigl(N_e+{1 \over g^2} n_m\bigr) 
\quad \hbox{to} \quad M=\vert Z\vert \quad \hbox{where} \quad 
Z=(an_e+a_dn_m)  \eqno(4.1)$$ 

where $n_e$ and $n_m$ denote the gauge-field and monople charges  
respectively, and  

the coefficients $a$ and $a_d$ will be explained later. 
\vskip 0.2truecm \noindent 8. 
It is symmetric with respect to a $Z_4$ symmetry which is the relic of the 
$R$-symmetry 

$(\theta_\alpha\rightarrow e^{i \epsilon}\theta_\alpha)$ 
that survives the axial anomaly breakdown. 
\vskip 0.2truecm \noindent 9. 
It has a holomorphic structure 
\vskip 0.2truecm \noindent 10. 
It has a duality that connects the weak and 
strong coupling regimes 
\vskip 0.2truecm \noindent 12. 
The duality generalizes to an $SL(2,Z)$ symmetry.  In section 6 we explain 
these 

last three concepts in a little more detail. 
\vskip 0.3truecm 
\noindent {\bf 5. }{\bf Spontaneous Symmetry-Breaking} 
\vskip 0.3truecm \noindent 
For $SU(2)$ this concept is very simple. From 
the form of the Higgs potential in (3.4) we see at once that there 
is a Higgs vacuum for $\phi=v\sigma$ where $v$ is any 
complex number and $\sigma$ is any fixed generator of $SU(2)$. 
Furthermore, for $v\not=0$ this 
breaks the gauge-symmetry from $SU(2)$ to $U(1)$. For other 
gauge-groups $G$ the corresponding statement is that ${\bf v}$ must 
lie in the Cartan subalgebra of $G$. On the other hand there is no 
spontaneous breakdown of supersymmetry. Thus the full breakdown is 
$$SU(2) \quad \rightarrow \quad U(1) \qquad N=2 \quad 
\hbox{supersymmetry unbroken} \eqno(5.1)$$ 
Indeed it is the fact 
that the supersymmetry is unbroken that gives the model its nice 
properties, since otherwise the classical properties would not be 
preserved after quantization. \vskip 0.2truecm \noindent After the 
spontaneous breakdown the restriction of the $N=1$ form of the 
classical Action (3.2) to the massless $U(1)$ fields takes the form  
$${\cal A}=\hbox{Im}\int d^4xd^2\theta_\alpha d^2\bar\theta_\beta \Bigl(\bar 
A A    \Bigr) +\tau_o\hbox{Im}\int d^4xd^2\theta_\alpha \Bigl( W_\alpha 
W_\alpha\Bigr)  \eqno(5.2)$$ 
Since the adjoint representation of $U(1)$ is trivial this 
Action is a free-field one.  However, in the quantum theory this 
does not mean that the effective Lagrangian is also free because, 
through the quantum fluctuations,  the massive fields induce 
interaction term for the massless ones. The first great virtue of 
the SW model is that these interactions have a very specific 
form. In fact they show that, due to the $N=2$ supersymmetry 
the {\it local} part of the effective Lagrangian can only be of the 
form 
$${\cal A}={1\over 2}\int d^4xd^2\theta d^2\bar\theta \Bigl(\bar A A_d-\bar A_d 
A\Bigr) +\hbox{Im}\int d^4xd^2\theta (\tau(A)) \Bigl( W_\alpha 
W_\alpha\Bigr)  \eqno(5.3)$$ where $$A_d=F'(A) \quad \hbox{and} 
\quad \tau(A)=F''(A)  \eqno(5.4)$$ 
for some function $F(A)$. Thus the effective Lagrangian is 
completely governed by the single function $F(A)$. Note that (5.3) 
is very similar to the classical Action which is the special case 
for which $F(A)={1 \over 2}\tau_o A^2$. As we shall see, the SW 
solution is actually a special Ansatz for the 
functional form of $F(A)$. 
\vskip 0.3truecm 
\noindent {\bf 6. }{\bf Holomorphy and Duality} 
\vskip 0.3truecm \noindent 
It is now easy to quantify what is meant by holomorphy and 
duality. Holomorphy is simply the statement that $F(A)$ depends only 
on $A$ and not on $\bar A$. Duality means that the physics described 
by the effective Action 
(5.3) is invariant with respect to the duality transformation.
$$\pmatrix {A\cr A_d)}\rightarrow \pmatrix {0&1 \cr 
-1 &0}\pmatrix {A\cr A_d)}  \qquad D_\alpha W_\alpha\rightarrow 
D_{\dot \alpha}W_{\dot \alpha}       \qquad 
\tau(A)\rightarrow (\tau(A))^{-1}  \eqno(6.1)$$ 
Note that the duality transformation is closely linked to the Legendre 
transform of $F(A)$ with respect $A$. By noting that in the free 
classical theory with $\theta_o=0$ the transformation (6.1) reduces to 
$$ \vec E\rightarrow \vec B \quad \hbox{and} \quad g\rightarrow {1 
\over g}   \eqno(6.2)$$ 
we see that it is just the generalization of 
well-known Maxwell-Dirac (MD) duality. Thus the Action (5.3) not only 
generalizes MD duality but puts it into a genuine dynamical model.  
Furthermore, the duality generalizes to 
$$\pmatrix {A\cr 
A_d}\rightarrow \pmatrix {p&q \cr r &s}\pmatrix {A\cr A_d}         
\quad \hbox{and} \quad \tau(A)\rightarrow {p\tau(A)+q \over 
r\tau(A)+s} \eqno(6.3)$$ 
where the matrix with entries $(p,q,r,s)$ 
is in $SL(2,Z)$. The integer-valuedness of the transformation 
follows from the requirement that, in the perturbation theory at 
least, it should change the $\theta$ angle only by multiples of 
$2\pi$ and should leave the mass-formula (4.1) form-invariant.
\vskip 0.3truecm\noindent 
{\bf 7. Perturbative and Non-Perturbative F(A)} 
\vskip 0.3truecm \noindent 
Before going on to describe the S-W Ansatz for F(A) we consider 
the perturbative contribution $F_p(A)$ of $F(A)$. This turns out 
to be $$ F_p(A)=\hbar A^2 
\hbox{ln}\Bigl(A^2/\Lambda^2\Bigr)    \eqno(7.1)$$ 
where $\Lambda$ is the renormalization scale. This is evidently a classical plus a 
one-loop expression but it is correct to all orders in perturbation 
because [3] the energy momentum tensor is in the  same multiplet as the 
axial current 
\beqq &\delta \Theta_{\mu\nu}=-{\bar \epsilon 
\over 4}\bigl(\sigma^{\mu\kappa}\partial_\kappa 
j^\mu+\sigma^{\nu\kappa}\partial_\kappa j^\mu\bigr)\nonumber  \\ \delta 
j_\mu^5 =i\bar \epsilon \gamma_5 j_\mu  \qquad & \delta 
j_\mu^\alpha=\epsilon \Bigl(2\gamma^\nu\Theta_{\mu\nu} 
-i\gamma_5\gamma^\nu\partial_\nu j_\mu^5 +{i \over 
2}\epsilon_{\mu\nu\kappa\lambda}\gamma^\nu\partial^\kappa 
j^\lambda_5\Bigr)\nonumber
\eeqq
\vspace{-7ex}
$$\eqno(7.2)$$
and in $N=2$ supersymmetry this situation is protected to all orders 
in perturbation. 
It follows that the quantum correction to the energy momentum tensor 
are similar to the axial anomaly, for which it is well-known that the 
one-loop result is exact to all orders. 
\vskip 0.2truecm \noindent 
For the non-perturbative part of F(A) the only solid a priori pieces of 
information are:
$$\hbox{Im} \bigl(F''(A)\bigr) \geq  0  \qquad  F_{np}(A)\not =0 
\qquad F_{np}(iA)=F_{np}(A) \eqno(7.3)  $$
and the fact that it vanishes for large $A$. 
The first relation from the convexity of the effective potential, specifically 
from the fact that $\hbox{Im}\bigl( F''(A)\bigr) $  is the coefficient 
of the kinetic term for the gauge-field, the second relation 
from 1-instanton computations and the third relation from the 
residue of $R$-invariance 
that is left after spontaneous symmetry-breaking. 
F(A) has to be guessed from this apparently meagre information. 
\vskip 0.6truecm \noindent 
\centerline{{\large The SW-Ansatz}}
\vskip 0.4truecm \noindent 
{\bf 8. Preliminaries}
\vskip 0.3truecm \noindent 
SW begin by reducing the problem to one in 
complex analysis by considering only the 
vacuum value $A=v$ of the chiral scalar superfield and 
determining the functional form of F(v). Afterwards, $(A)$ can be 
recovered by the simple substitution 
$F(v)\rightarrow F(v +\tilde A)\equiv F(A)$. 
This is analogous to the substitution 
$$V(m,f,g)  \quad \rightarrow \quad V(m+f\phi+g\phi^2,f+g\phi, 
g)=V_{eff}(\phi) \eqno(8.1)$$
which is made to obtain the effective potential from the partition  
function P(m,f,g) of a standard renormalizable theory with a single 
scar field $\phi$ with masses m, and coupling constants f and g. 

\noindent Next, they note that asymptotic freedom allows them to 
identify the perturbative  region  
as the large scale one $v\rightarrow \infty$  and thus
$$ \tau(v) \rightarrow  {i \over \pi} 
\hbox{ln}(v)  \quad \hbox{for} \quad 
v\rightarrow \infty \eqno(8.2) $$ 
Since v is expected to be singular in the small scale (strong 
coupling) region one also postulates the existence  of a 
universal (complex) parameter $ u\in C$ normalized so that 
$a(u)\rightarrow u^2$ for $ u\rightarrow \infty$.
Assembling all this information they reduce the problem to the 
search for a function  $\tau(u)$ such that
$$\hbox{Im}\bigl(\tau(u)\bigr)\geq 0 \quad \hbox{and} \quad 
\tau(u)_{u\rightarrow \infty }\rightarrow {i\over 2\pi 
}\hbox{ln}(u) \eqno(8.3)$$
The procedure for choosing a $\tau(u)$ to satisfy (8.3) 
is actually rather similar that used by Veneziano in choosing  
his formula for the S-matrix S(s,t,u), where s,t and u are the 
invariant squares of the momenta. That is to say, instead of 
computing the function directly from the underlying theory  
one uses its properties (symmetries, boundary conditions 
etc.), to try to guess what it should be. Indeed duality  
(actually triality) plays here a role which is analogous to that 
played by crossing symmetry (symmetry in s,t and u) in the Veneziano case.
But first one has to decide the general class of functions out 
of which the function $\tau(u)$ should be chosen. The standard 
class of functions which map the upper part $C_+$ of the complex 
plane into itself modulo subgroups of $SL(2,Z)$ is the class  
of {\it Fuchsian} functions [5]  
and the choice will be made  out of these. So, to put the 
results in perspective, we digress for a moment to consider  
Fuchsiam functions or maps. 
\vskip 0.3truecm \noindent 
{\bf 9. Fuchsian Maps}
\vskip 0.3truecm \noindent 
The Fuchsian maps $\tau(u)$ are maps from $C_+$ to $D=C_+/G$ where $G$ 
is a subgroup of $SL(2,R)$ (restricted in our case to $SL(2,Z)$. 
Thus they map $C_+$ into 
fundamental domains $D \in C_+$ whose $G$-equivalent copies fill 
out $C_+$. The domains $D$ are circular polygons (ones whose sides 
may be straight lines or circles) and in our case will stretch out 
to infinity. In general the 
polygons may not be of genus zero because the  sides may be 
have to be identified in a non-trivial manner. The corners of the 
polygon correspond to points on the real axis in the $u$-variable 
and if the genus of the polygon is zero $u$ can be 
extended to cover the whole complex plane, which 
compactifies to the Riemann sphere. Otherwise the 
compactification of the $u$-space leads to a Riemann surface of higher genus. 
The essential point is that these maps guarantee that 
Im$(\tau(u))\geq 0$. 
\vskip 0.3truecm \noindent 
{\bf 9. The Schwarzian Derivatives }
\vskip 0.3truecm \noindent 
The Fuchsian functions $\tau(u)$ are apt to be complicated 
but a great simplification is achieved by considering not the 
functions themselves but their Schwarzian derivatives 
$$S(\tau)={\tau''' \over \tau'}-{3 \over 2} \Bigl( {\tau'' \over 
\tau' } \Bigr)^2 \eqno(9.1)$$
In general the main property of Schwarzian derivatives is that they are 
invariant with respect to the modular transformations (6.1). 
However, in the case of Fuchsian functions they have the added 
advantage that they are simple meromorphic functions of the form 
$$S(\tau(u))=\sum_{i=1}^{i=n}\Bigl\{  {1 \over 2} {(1 -\alpha_i^2) \over (u-a_i)^2 } 
+ {\beta_i \over (u-a_i)}\Bigr\} \qquad S\bigl(\tau(u)\bigr)\rightarrow{1 \over 
u^2} \quad u\rightarrow \infty \eqno(9.2)$$
and this is why it is convenient to use $S(\tau)$ rather than $\tau$ 
itself. One $S(\tau)$ is known there is a simple and elegant way to 
recover $\tau$ from it, namely to write 
$$\tau={y_1 \over y_2} 
\quad \hbox{where} \quad y''+{1 \over 2}S(\tau(u))y=0     
\eqno(9.3)$$
Such functions $\tau$ are tailor-made for the S-W model where
$\tau(u)=a'_d/a'$. All but the $\beta$-parameters in (9.2) have a simple 
geometrical meaning. The number $n$ is the number of corners on the 
polygon $D$ (excluding the point at infinity) and the 
parameters $a_i$ and $\alpha_i$ are the locations of the corners 
and the internal angles of the polygon respectively. Because of the 
freedom of choosing the axes in $u$-space the number of 
independent parameters in $Q(u)$ is $3n-2$. The boundary condition 
puts two further restrictions on the on the $\beta$'s and this 
reduces the number to $3n-4$. This already shows that $n\geq 2$.  
It is clear that the real choice in choosing a Fuchsian function is to 
choose $n$ and then choose the $3n-4$ parameters in the meromorphic 
function Q(u). 
\vskip 0.3truecm \noindent 
{\bf 10. S-W Choice}
\vskip 0.3truecm \noindent 
The question is: which Fuchsian function to choose? 
The S-W choice is made by adding two further inputs 
to the basic conditions, namely   
\vskip 0.2truecm \noindent 
1. Minimality: \hskip 1truecm  $n=2$
\vskip 0.2truecm \noindent
2. Duality: \hskip 0.4truecm  
$M_{\infty}=\pmatrix{1&2\cr0&1}\quad \leftrightarrow  
\quad M_1=\pmatrix{1&0\cr2&1} $
\vskip 0.2truecm \noindent
Duality in this sense means that the monodromy matrix $M_1$ at  
one of the singularities is the dual of the monodromy matrix 
$M_{\infty}$ at infinity. 
Physically, it means that the asymptotic freedom of g for  
large scales is the dual of the infra-red slavery of g  
(asymptotic freedom of $g^{-1})$ for low scales. It turns 
out that $M_{\infty}$  and $M_1$  generate a monodromy  
group $\Gamma_2$ which includes the monodromy  group of the second 
singularity, and consists of all matrices of the form 
$$I+2\pmatrix{e&f\cr g&h} \quad e,f,g,h \in Z 
\eqno(10.1)$$
According to our previous analysis this determines a unique 
Fuchsian map (modulo the positions of the two singularities which 
are normalized to be $a_i=\pm1$) and it turns out to be the map with 
$\alpha_i=0$ and $\beta_i=\pm1/4$. Thus with SW Ansatz equation (9.3) 
becomes  
$$\tau(u)= {y_1 \over y_2} \quad \hbox{where} \quad 
y''+{1 \over 4}\Bigl[{3+u^2 \over (u^2-1)^2}\Bigr]y=0 \eqno(10.2)$$
There is actually a simplification in this case on changing to the 
variable $a$, where $a'=y$ because then (10.2) becomes 
$$\tau={a'_1 \over a'_2} \quad \hbox{where} \quad 
a''+{1 \over 4} \Bigl({1 \over u^2-1}\Bigr) a=0  \eqno(10.3)$$
and the differential equation in (10.3) is just a hypergeometric 
equation. So finally $ \tau(u)$  is simply the ratio 
of the derivatives of two simple hypergeometric functions. 
In fact the function $\tau^{-1}(u)$  is a well-known 
automorphic function called the elliptic modular function. 
Thus when all the smoke has cleared away it turns out that the SW Ansatz 
is to propose that $F(A)$ is such that $F''(A)$ is the inverse of 
the elliptic modular function! 
\vskip 0.3truecm \noindent 
{\bf 11. Uniqueness}
\vskip 0.2truecm \noindent 
The S-W Ansatz is certainly not the only Ansatz compatible with the basic 
conditions
$$\hbox{Im}(\tau(u))\geq 0 \qquad 
\tau(u)\rightarrow {i \over 2\pi}\hbox{ln}(u)  \eqno(11.1)$$
Indeed, even for two singularities, the Schwarzian could be 
$$S(\tau(u))={(1-\alpha^2)\over 2(u-1)^2}+{(1-\gamma^2) \over 
(u+1)^2}-{1 \over 2} {(2-\alpha^2-\gamma^2)\over (1-u^2)}\eqno(11.2)$$
where $\alpha, \gamma=0, {1 \over 3},{1 \over 2}$, with monodromy 
groups called  $\Gamma_2$, $G_2$ 
and $G_\theta$. The solution with $G_\theta$ is ruled 
out because it is not $R$-invariant 
but $G_2$ remains as a reflexion-invariant alternative. For more than 
two singularities 
there are many more possibilities. For example a three-singularity 
reflexion-invariant solution with monodromy group $SL(2,Z)/Z_2$ is provided by 
$\tau (u)=J(u^2)$,  where $J$ is the standard modular function. 
\vskip 0.3truecm\noindent 
What distinguishes the S-B solution is that it 
carries the dual symmetry that was used by S-W 
as input. 
The point is that $\Gamma_2$ is an {\it invariant} subgroup of SL(2,Z) 
and
$$\hbox{SL(2,Z)}/ \Gamma_2 =P_3    \eqno(11.3)$$
where $P_3$ is the permutation group of order 3. 
Mathematically the permutation group $P_3$ interchanges the point at 
infinity and the two singularities and 
physically it interchanges gauge-fields, monopoles and dyons. The 
original duality input emerges as the symmetry between the 
the gauge-field at infinity and the monople at one of the 
singularities. 
\vskip 0.3truecm\noindent 
{\bf 12. Correctness}
\vskip 0.3truecm\noindent 
Since the S-W Ansatz is not unique one has to check whether 
it is, in fact the correct choice. In principle this can be done 
by making direct computations of the non-pertubative part of the 
Action using instanton computations. In practice this has been done 
[2] only for arbitrary gauge-groups in the 1-instanton configurations and 
for $SU(2)$ in the 2-instanton configurations and in these cases 
the SW Ansatz agrees with direct computations. 

\noindent The general idea of these computations is as follows: 
For any Fuchsian function 
satisfying the boundary conditions and R-invariance we have the 
asymptotic expansion 
$$F(v)=v^2+\hbar v^2\hbox{ln}\bigl({v \over \Lambda}\bigr)^2 
+v^2\sum c_m \Bigl({\Lambda \over v}\Bigr)^{4m}    \eqno(12.1)$$
In the direct instanton computations $F(v)$ is supposed to be  
the partition function and the contributions of the various powers 
in $m$ are supposed to come from the instanton sectors of 
topological charge m. The idea, therefore, is to compute 
the partition function in an m-instanton background under the 
assumption that the scalar field has a non-zero value v on the 
sphere at infinity. What one does in practice is to first choose a 
background for the gauge and fermion fields by the conditions 
$$F_{\mu\nu}=F^*_{\mu\nu}  \qquad \gamma^\mu D_\mu \psi=0 
\eqno(12.2)$$
where $F^*$ denotes the dual of $F$. These fields are parametrized 
by the $8m$ ADHM parameters $\rho,\nu$ for self-dual gauge-fields, 
where $ \rho$, in particular, denotes the size of the instanton, 
and the 8m parameters $\eta, \bar \eta$ for their zero-mode fermion fields. 
One then postulates that the scalar field is given by 
$$D^2\phi=[\bar \psi,\psi]  \qquad \phi(\infty)=v \eqno(12.3)$$
and is thus a functional of these $16n$ parameters. Finally one 
postulates that long-range part of the partition function comes only from 
the surface term tr($\int \phi^\dagger D_r\phi)$ in the Action 
(3.4) and that the short range part drops out because the bosonic 
and fermionic contributions cancel on account of supersymmetry. In 
that case the partition function evidently takes the form  
$$P(v)=\Lambda^{4n} e^{{8n\pi^2 \over g^2}}
\int d\rho d(\nu)\rho^{4n-3}d\rho d\nu d\bar \eta d\eta  
e^{\int d\vec \Omega.(\phi,\vec D\phi)} \eqno(12.4)$$
For dimensional reasons these integrals are of the form 
$$P(v)=\Lambda^{4m} e^{{8m\pi^2 \over g^2}}
\int \rho^{4n-3}d\rho d(\nu) e^{-(a\rho)^2f(\nu)}=
a^2\Bigl({\Lambda\over v}\Bigr)^{4m} e^{{8m\pi^2 \over g^2}}e_m  
\eqno(12.5)$$
where the $e_m$'s are dimensionless constants. It is these constants 
that are to be compared with the SW constants $c_m$ and, as 
mentioned earlier, the coefficients computed up to now are in agreement.  
So the instanton computations provide reasonably strong  support for 
the correctness of the SW Ansatz. As a check 
on the sensitivity of the result we have computed $c_1$ for the 
alternative $n=2$ Ansatz and it is different from the SW one.

\noindent It must be admitted that the validity of the instanton 
computations as described above is not quite clear, because the 
equations (12.2) and (12.3) are regarded only as approximations and 
the background configurations described by them are neither 
solutions of the classical field equations nor 
supersymmetric-invariant. However, we hope to present a more convincing 
argument for the validity of the instanton  computations in a later paper. 
\vskip 0.5truecm \eject\noindent 
{\bf References} 
\vskip 0.2truecm \noindent 
[1] N. Seiberg and E. Witten, Nucl. Phys. {\bf B426} (1994) 19 {\bf 
430} (1994) 485(E)
\vskip 0.2truecm \noindent 
[2] D. Finnell and P. Pouliot, Nucl. Phys. {\bf B453} (1995)
 225
\vskip 0.1truecm \hskip 0.01truecm 
N. Dorey et al. Phys. Rev. {\bf D54} (1996) 2921; hep-th/9607202
\vskip 0.1truecm \hskip 0.01truecm  
K.Ito and N. Sasakura, Phys. Lett. {\bf B382} (1996) 95
\vskip 0.1truecm \hskip 0.01truecm 
A. Yung, hep-th/9605096 
\vskip 0.1truecm \hskip 0.01truecm 
F. Fucito and G. Travaglini, hep-th/9605215 
\vskip 0.1truecm \hskip 0.01truecm 
T. Harano and M. Sato, hep-th/9608060
\vskip 0.2truecm \noindent 
[3] J. Wess and R. Bagger, {\it Supersymmetry}, Princeton University Press 
(19.)
\vskip 0.1truecm \hskip 0.01truecm 
P.West, {\it Introduction to Supersymmetry and Supergravity} World 
Scientific (1990) 
\vskip 0.2truecm \noindent 
[4] C. Montonen and D. Olive, Phys. Lett. {\bf B72} (1977) 117
\vskip 0.2truecm \noindent 
[5] A. Erdelyi et al. {\it Higher transcendental functions}, 
Vol. $3$, McGraw-Hill, NY (1955)
\vskip 0.1truecm \hskip 0.01truecm  
Z. Nehari, {\it Conformal Mapping}, McGraw-Hill, NY (1952)
\vskip 0.1truecm \hskip 0.01truecm 
R.C. Gunning, {\it Lectures on Modular Forms}, 
Princeton University Press (1962)  
\vskip 0.1truecm \hskip 0.01truecm 
L. R. Ford, {\it Automorphic Functions}, Chelsea 
Publishing Company, NY (1951)

\end{document}